\documentstyle[prl,aps,epsf,multicol]{revtex}
\begin{document}
\def\be{\begin{equation}}
\def\ee{\end{equation}}
\def\bearr{\begin{eqnarray}}
\def\eearr{\end{eqnarray}}
\def\tc{$T_c~$}
\def\tcl{$T_c^{1*}~$}
\def\c2{ CuO$_2~$}
\def\ruo{ RuO$_2~$}
\def\lsco{LSCO~}
\def\bi{Bi-2201~}
\def\tl{Tl-2201~}
\def\hg{Hg-1201~}
\def\sro{$Sr_2 Ru O_4$~}
\def\rc{$RuSr_2Gd Cu_2 O_8$~}
\title{Spinon deconfinement above a finite energy gap in\\ 
2d Quantum Heisenberg Antiferromagnets}
 
\author{ G. Baskaran \\
Institute of Mathematical Sciences\\
C.I.T. Campus
Madras 600 113, India }

\maketitle

\begin{abstract}
The familiar spin-$\frac{1}{2}$ quantum Heisenberg antiferromanget in a 2d 
square lattice is shown, within the non linear sigma model approximations, to 
be another novel state of matter that has excitations with fractional quantum 
numbers above a finite energy gap.  The 1-skyrmion with an energy 
$\approx 2\pi J$ 
is shown to be made of two `deconfined spinons' or `SU(2) vortices'. 
The many skyrmion operator and the wave functions that we have found
are strikingly similar to quantum Hall quasi particle 
operators and wave functions. We also predict the presence of finite 
energy `spin-S spinon' for a general spin-S Heisenberg antiferromagnets 
in 2d.  Some consequences are briefly discussed. 
 
\end{abstract}

\begin{multicols}{2}[]
Neutral spin-$1\over2$ fermionic excitations\cite{pwascience}, now called 
`spinons'\cite{ABHZ} were conjectured to be present in the quantum 
spin liquid states of spin-half 2d Heisenberg antiferromagnets, 
by Anderson in 1987. While this was readily shown to be
present in the RVB mean field analysis\cite{bza} of the spin liquid
vacua and short range RVB states\cite{krs}, their presence as a 
finite energy deconfined excitation in the ordered antiferromagnetic 
vacua has remained unclear.  

A conjecture by Dzhyaloshinskii, Wiegman and Polyakov\cite{dwp} 
and the early work of Wilczek and Zee\cite{dwp}
suggested statistics transmutation of skyrmion 
in ordered spin-$1\over2$ Heisenberg antiferromagnets in 2d. 
Anderson, John, Doucot, Liang and the present 
author\cite{AJDLB} however conjectured that finite energy 
`half-skyrmion' or `meron'\cite{gross} are the `deconfined' 
spinons based on some heuristic arguments. 

Recent inelastic neutron scattering results covering a large 
energy and momentum range in the insulating cuprates\cite{neutron}, 
and also Raman\cite{raman} and infrared
measurements also makes the search for any signatures 
of spinons at low and high energies very meaningful and urgent.

The aim of the present letter is to study the spectrum of quantum 
Heisenberg antiferromagnet in 2d and look for deconfined spinons
above a finite energy gap, within the O(3) non linear sigma (NLS)
model approach.  We look carefully at Belavin Polyakov\cite{belavin} 
n-skyrmion {\em static classical solutions}. The mathematical structure 
of the n-skyrmion solution in one particular parameterization readily 
suggests that each skyrmion is made of 2n `constituent  point 
particles\cite{gross,fateev,din}.

To understand the quantum dynamics of these  constituent particles we 
construct the skyrmion operator for our spin-$1\over2$ Heisenberg
antiferromagnet and discover 
that the creation operator for these constituent particles have 
mathematical structure similar to Laughlin's quasi hole and quasi 
particle operators of quantum Hall effect. By a Berry phase analysis
we show their spin to be half. Asymptotic form of the modulus of 
the n-skyrmion wave function in terms of collective coordinates 
is also found. 

Our starting point is the spin-S quantum Heisenberg antiferromagnet 
in a 2d square lattice with nearest neighbor interactions. 
Following the standard  derivation\cite{haldane1} one arrives at 
the O(3) NLS 
model action along with the important lattice sum of Berry phases:  
\be
S = \frac{\rho^{}_0}{2}\int dx dy~dt \bigl( 
[{\bf \partial_{\mu}}{\bf n}({\bf r})]^2
- \frac{1}{v_s^2}[{\partial_t\bf n}({\bf r})]^2\bigr)
+ i S^{}_B[{\bf n}]
\ee
Here $\mu = x,y$ and ${\bf n}({\bf r})$ is a normalized 
(${\bf n}({\bf r})\cdot {\bf n}({\bf r}) = 1$) sub lattice magnetization 
vector. 

The coefficient $\rho^{}_0 \approx J$ for $S = \frac{1}{2}$ case,
and $v_s$ is the spin wave velocity. The Berry phase term  is a 
lattice sum 
\be
S^{}_B[{\bf n}] = 2\pi S \sum_{m,n} (-1)^{|m| + |n|}_{} \Omega^o_{}
({\bf n}_{m,n})
\ee
Here the integers (m,n) stand for a lattice site and 
$ \Omega^o_{} ({\bf n}^{}_{m,n}) = \int \frac{dt~du}{8\pi} 
{\bf n}^{}_{m.n}\cdot( \partial^{}_t{\bf n}^{}_{m.n}\times 
\partial^{}_u{\bf n}^{}_{m.n})$ 
is the single site Berry phase. 

Inspired by the conjecture of reference (6) we look at the finite energy
topological solutions in our search for spinons.  The multi skyrmion 
solutions found by Belavin and Polyakov\cite{belavin}
are extended objects with non trivial topology.  Their Eucledian 
solutions of the  O(3) model in  (1 + 1) d become the time independent 
classical solutions of our O(3) model in (2 + 1) d.

An n-skyrmion solution is given by
\be
w(z) = \prod_{i = 1}^{n} \Bigl(\frac{ z - a^{}_i}{ z - b^{}_i}\Bigr)
\ee
Here $z = x + i y$. The n complex co-ordinates $a^{}_i$ and 
$b^{}_i$ characterize the skyrmion solution. 
The function $w(z)$ and the sub lattice magnetization
${\bf n}({\bf r}) \equiv  (\sin \phi({\bf r}) \cos \theta({\bf r}),
\sin \phi({\bf r}) \sin \theta({\bf r}), \cos \phi({\bf r}))$
are related by
\begin{equation}
 w(z) \equiv \cot \frac{\phi({\bf r})}{2} e^{i\theta({\bf r})}
\ee
The n-antiskyrmion solutions are obtained by replacing $z$ by ${\bar z}$.
In our convention the spins at infinity $\bf{n}(\infty) = (1,0,0)$
for the ground as well as the skyrmion/antiskyrmion states, 
since $w(\infty) = 1$.

The energy of the n-skyrmion solution is given by
\be
E^{n}_{cl} = 4\pi\rho_0 n,
\ee
a constant independent of the skyrmion coordinates $\{a^{}_i,b^{}_i\}$ !  
This means that at the classical level the n-skyrmions do not 
interact; thus the set $\{a^{}_i,b^{}_i\}$ represents
the 2n `zero mode' (2 dimensional) co-ordinates of the n-skyrmion. 
We will call the a and b coordinates of the 2n constituent 
particles of an n-skyrmion.
    
Physically the co-ordinates a and b represent the `centers' 
of local vortex distortions of the xy compounts of 
the 3 vector field ${\bf n}$. 
However, unlike the $\pm$ vortex pair
of an xy model, there is a freedom in the choice of the positions
of the vortex pair; for a 1-skyrmion any two diametrically opposite 
points on the circle $|z -\frac{a+b}{2}| = |\frac{a-b}{2}|$ can 
be chosen as centers of the vortex pair, 
with the local vortex distortion occurring in appropriately rotated 
2d plane in the spin space.  It is remarkable that the extended 
spin twists of a 1-skyrmion, which is more of a`ring like' distortion 
of the ${\bf n}$ field, than two point vortices, is characterized by just
two points a and b, which we will later elevate to the level of 
particle degrees of freedom in the plane. 
We say that a n-skyrmion is made of 2n `SU(2) vortices'. 

It should be remarked that the Belavin Polyakov's exact solutions 
describe either skyrmions or antiskyrmions. No exact solution
containing both skyrmions and antiskyrmions exist. In approximate 
solutions containing 1-skyrmion and 1-antiskyrmion there is a dipole 
like interaction between them even at the classical level\cite{gross}.

The two SU(2) vortices of a 1-skyrmion state are finite energy 
solutions and they have twisted their way out of the $\log |a-b|$ 
energy dependence, characteristic of xy vortices in 2d.
We will see later that quantum fluctuations produce repulsion between  
the two SU(2) vortices of a 1-skyrmion ! Further, once an 
xy anisotropy is introduced the SU(2) vortices degenerate to two 
xy vortices with an energy diverging as 
$ \frac{J_{x} - J_{z}}{J_z} \log|a - b|$ as $|a-b| \rightarrow \infty$. 

An n-skyrmion/antiskyrmion carries a topological `quantum number' 
or a `winding number' $\pm~ n$, the degree of the map
$(x,y) \rightarrow S^2$ of the field ${\bf n}(x,y)$. 
Here $S^2$ is the order parameter space.The `topological density' 
$q({\bf r})$ is given by   
\bearr
q ({\bf r}) & = & 
\frac{1}{8\pi}{\bf n}^{}_{}({\bf r})\cdot 
(\partial^{}_x {\bf n}^{}_{}({\bf r})\times
\partial^{}_y {\bf n}^{}_{}({\bf r})) \\
& = & \frac {(
\partial_z {\bar w}(z) \partial_{\bar z} w({\bar z}) - 
\partial_z w({\bar z}) \partial_{\bar z} w(z))} 
{\pi( 1 + |w(z)|^2 )^2} \nonumber
\eearr
And $\int dx dy ~ q({\bf r}) = n$ is the winding number.

To keep comparison with other formalisms, the above topological 
density is related to the U(1) magnetic flux of the RVB gauge 
field\cite{gbgauge}:
\be
{\bf S}_i \cdot ({\bf S}_j \times {\bf S}_k) \equiv
  i(\tau_{ij} \tau_{jk} \tau_{ki} - 
  \tau_{ik} \tau_{kj} \tau_{ji})
    \sim e^{i \oint {\bf A}_{rvb}\cdot{d\bf l}}
\ee
where $\tau_{ij}  = \sum_\sigma c^\dagger_{i\sigma} c^{}_{j\sigma}$
and $n^{}_{i\uparrow}+n^{}_{i\downarrow} = 1$

To understand the quantum dynamics of the skyrmion excitation and 
its constituent particles for our spin-$\frac{1}{2}$ 
Heisenberg model on the square lattice, we first construct the 
operator for the 1-skyrmion on the lattice. (No exact classical
solution is available for the `lattice version' of the NLS model
in 2d; when $|a-b| >> 1$, in units of the lattice parameter,
the continuum skyrmion solution should be a good 
approximation to the lattice solution). The spin wave ground state 
of our spin-$\frac{1}{2}$ Heisenberg antiferromagnet is
\be
\Psi_{sw} [z,{\bar z}] \sim (-1)^{N_A^{}} 
\exp \{ -\sum_{i < j} f(|z_i - z_j|)\} 
\ee
where $z_i \equiv m_i + i~ n_i $ stand for the lattice co-ordinates of 
the $N^{}_0\over2$ down spin electrons in the square lattice containing 
$N^{}_0$ sites. 
This is a {\em hard core
boson} (or equivalently $s^z_{}$) representation of the spin wave ground 
state, where the 
Jastrow factor $f(|z_i - z_j|) \sim {\frac{1}{|z_i - z_j|}}$ arise from 
the zero point quantum fluctuations of the gapless spin wave modes of 
the ordered antiferromagnet.  And  $(-1)^{N_A}_{}$ is the Marshall sign 
factor,  where $N_A^{}$ counts the number of down spins on sub lattice A. 
The ODLRO exhibited by the above ${N^{}_0\over2}$ hard core boson 
wave function represents the long range antiferromagnetic order,
with a finite sub lattice magnetization ${\bf m}$, lying in 
the xy plane. We can fix the  direction of ${\bf m}$ in 
the xy plane by  giving an additional coherent superposition with 
respect to $S^z_{total}$ (the total number of hard core bosons) 
in the above wave function\cite{note1}.

After some algebra the co-ordinate representation of the 1-skyrmion 
operator is shown to be 
\be
\hat {\bf O}^{(1)}_{Sk}(a,b) \equiv \prod_{i = 1}^{\frac{N^{}_0}{2}} 
\Bigl(\frac{ z_i - a}{z_i - b} \Bigr).
\ee
While $w(z)$ of equation (3) represents the distortion of the sub lattice 
magnetization in space, equation (9) provides the operator that 
produces the distortion. To prove the above, it is convenient to put 
the sub lattice magnetization along the + x-axis in the ground state
and use the $s^z$ representation, $  \hat {\bf O}^{(1)}_{Sk}(a,b) 
\equiv \prod_{i = 1}^{N^{}_0} \Bigl(\frac{ z_i - a}{z_i - b} 
\Bigr)^{s^z_i + {1\over2}}$. (Note, unlike equation (9), here 
$z_i$'s run over all sites !).

This is an important result, which shows that a {\bf 1-skyrmion operator
actually creates one Laughlin\cite{laughlin}
 quasi-hole and one quasi-electron like 
object}, when acting on the hard core  bose fluid. 
Similarly an n-skyrmion state can be created by a product 
of n of Laughlin quasi-holes and n of quasi electron like operators.
The operator for anti skyrmion is obtained by replacing $\{z_i\}$ by
$\{\bar z_i\}$ in equation (9). Our quasi hole operator exactly 
coincides with Kalmayer-Laughlin's spinon operator\cite{laughlin}. 

Since at the classical level, the a and b co-ordinates of the n-skyrmion
solution represent the zero-mode co-ordinates, it is legitimate to elevate
them to the level of new low energy collective co-ordinates or `particle  
degrees of freedom\cite{din}, as in the case of Laughlin's quasi holes and 
quasi electrons.

While the a and b `quasi particles' (SU(2) vortices) do not have 
any interaction at a classical
level, the modified quantum fluctuations in the the presence of a skyrmion can 
induce interactions among the constituent particles. The induced interaction
$V_{qu}(a,b)$ is obtained from the 
difference in zero point energy of the spin wave modes in the presence
$ \hbar\omega_{\mu}^{(a,b)}$  and absence $ \hbar\omega_{\alpha}^0$  
of a 1-skyrmion:
\be
V_{qu}^{}(a,b) \equiv \frac{1}{2}\sum_{\mu}\hbar 
\omega^{(a,b)}_{\mu} - \frac{1}{2}\sum_{\alpha}\hbar 
\omega_{\alpha}^0
\ee
Using the method of Rodriguez\cite{rodriguez}and also Marino\cite{marino}, 
who did not use the constituent particle 
interpretation, we have calculated the induced interaction between 
an a and b particle. We find that the a and b particles always repel. 
It is energetically advantageous 
for the constituent particles to be infinitely apart and reduce
the zero point fluctuation energy. In particular the quantum fluctuation
corrected energy of a 1-skyrmion state is:
\bearr
E_{qu}^{n=1} & (|a-b| \rightarrow 0) & \approx  4 \pi J \nonumber \\ 
E_{qu}^{n=1} & (|a-b|\rightarrow \infty) & \approx  2 \pi J \nonumber 
\eearr
This proves that the two constituent particles are indeed deconfined.

I have obtained an asymptotically exact {\em modulus of 
the wave function} of the `ground state' of the n-skyrmion state, 
for large separation of the a and b coordinates as 
\be
|\Psi_n^{}[a,b]| \sim \prod_{i < j}|a_i-a_j| |b_i - b_j| 
\prod_{i,j} |a_i - b_j|^{-1}
\ee
The method I have devised involves expressing the path integral for 
vacuum to vacuum
amplitude in a way that gives the desired wave function in terms of 
the Jacobian of transformation from the function $w(z)$ to the zero
mode coordinates. A plasma analogy shows that
the a and b particles are indeed unbound, giving another proof for 
the deconfinement of the a and b particles in the skyrmion state. 

The spin of a skyrmion or the constituent SU(2) vortices and their 
exchange statistics are rather ill defined in view of the large 
quantum fluctuation contained in the spin wave ground state. 
The `mathematically sharp' point particles a and b are extended
physical objects with power law form factors. (This is similar to the 
case\cite{haldane2} of vortices in 2d superfluid He 4). The integrated
missing or excess density around a vortex gives the z-component of
the spin of the a and b particles. This calculation turns out to 
be hard with our hard core bose fluid. So we take a Berry phase
approach. 

In our Berry phase analysis the spin of the constituent particle
or the SU(2) vortices appears as a `chain anomaly'. We consider 
a one skyrmion solution and analyze the lattice sum of the Berry 
phases under a global rotation of all the spins through $ 2\pi $
about x-axis, the direction of sub lattice magnetization at infinity.
If the spin projections
of the constituent particle along the x-axis are $\sigma_a $ and
$\sigma_b$, the corresponding lattice sum of Berry phases should 
contribute $2\pi\sigma_a$ and  $2\pi\sigma_b$ for the a and b 
particles in a spatially separated fashion.

\begin{figure}[h]
\epsfxsize 6cm
\centerline {\epsfbox{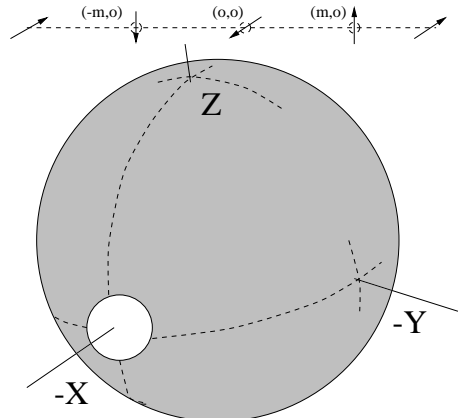}}
\caption{ The `hole' in $S^2_{}$ of the map 
$ (x,t) \rightarrow S^2$ of the \\
field {\bf n}(x,n;t) for a $n \neq 0$ chain. 
The `hole' just vanishes\\ for the n = 0 chain;
inset shows its ${\bf n}(x,0;t)$ at t = 0.} 
\end{figure}

We find an interesting phenomenon which we call a `chain anomaly'.
We group, following Haldane\cite{haldane1}, the square lattice
sum of Berry phases into sums over chains, running parallel 
to the x-axis:
\be
S_B =  2\pi S \sum_n (-1)^{n} \Omega^{1d}_n 
\ee
where the n-th chain Berry phase is given by the sum
\bearr
\Omega^{1d}_n & \equiv & \sum_{m = -N, .. N-1} (-)^m 
\Omega^o_{}({\bf n}_{m,n}) \nonumber \\
& = & \int \frac{dx~dt}{8\pi} {\bf n}^{}_{}(x, n;t)\cdot (\partial^{}_x 
{\bf n}^{}_{}(x,n;t)\times
\partial^{}_t {\bf n}^{}_{}(x,n;t))
\eearr
The chain Berry phase counts the winding number of map
$(x,t) \rightarrow S^2$ of the field  $ {\bf n}(x,n;t)$ 
for n-th chain.

We choose the length of our lattice along the x-axis to be 2N, 
where N is an odd integer.
For convenience we choose $(0,0)$ to be the center of the 
square lattice and the two skyrmion co-ordinates  
to be the lattice sites, $ a = (m,o)$ and $ b = (-m,0)$ on the 
x-axis at time t = 0 and T. Our global rotation amounts to giving 
the following time dependence $a(t) = m e^{i\frac{2\pi t}{T}}$ and 
$b(t) = -m e^{i\frac{2\pi t}{T}}$ to the constituent particle
co-ordinates.

During the time evolution, the spin field
${\bf n}(x,n;t)$ of the 1-skyrmion state of any chain n 
attempts to wrap the $S^2$ sphere.
All, except the x-axis chain ( n = 0) fail to wrap the 
$S^2$ sphere completely. They all leave a hole, as shown 
in figure 1, making the winding number (Berry phase) 
identically zero. The x-axis chain 
just manages to wrap the $S^2$ sphere, once for the chain sum
$-N, .. -1$ and again in the opposite sense for the chain sum
$0,1,..N-1$, thereby  contributing a phase of $\pi$ for the left half
and $-\pi$ for the right half of the chain. We identify these
two phases, which arise predominantly from the a-vortex region
and b-vortex region,  with the spin-$\frac{1}{2}$ Berry phases 
of a and b particles.

The lattice sum of Berry phases can be grouped in many different
ways and {\em we always find one singular chain}
passing through two diametrically opposite points of the 
circle $|z -\frac{a+b}{2}| = |\frac{a-b}{2}|$, of our 1-skyrmion
solution. The two ends of the chain either close on themselves 
or go off to infinity. The spin field ${\bf n}$
for these singular chains trace a closed curve in $S^2$ 
connecting two antipodal points at any given time; {\em it 
undergoes `$\pi$-twist' twice, corresponding to the presence of 
two spinons}. 
For example, the circle $|z -\frac{a+b}{2}| = |\frac{a-b}{2}|$
itself is a singular chain, for which the ${\bf n}$ field traces
a great circle on $S^2$ at any given time. 

In spin-$\frac{1}{2}$ Heisenberg chain each $\pi-twist$ 
of ${\bf n}$ field in space corresponds to one spinon\cite{fadeev}.
In a sense, {\em in the 1-skyrmion configuration a `1d-chain' 
or a `string' containing two spinons is embedded in the 2d plane 
in an irreducible and dynamical fashion}. 

Our chain anomaly is missed if we use Haldane's 
argument\cite{haldane1}
for the calculation of the Berry phase.  Haldane suggested that 
if the ${\bf n}(x,y;t)$ field is continuous and non singular 
(as it is for the case for n-skyrmion solution), the chain Berry 
phases should be all identical and as a result the staggered sum 
should be identically zero, for an even number of chains. However, 
we find that our singular chain containing the a and b co-ordinate 
is an exception to this and it contributes a Berry phase of $\pi$ 
and $-\pi$ to the two particles a and b.

The two SU(2) vortices or the two spinons of a 1-skyrmion state, 
by the argument given above, carry  spin half projections of value 
$1\over2$ and $-{1\over2}$ along the x-axis. In addition the spinons 
carry $ +1$ chirality quantum number because they are constituent
particles of a + 1-skyrmion state. Thus a spinon 
in addition to a spin quantum number carries a $\pm 1$ 
chirality quantum number. As the antiferromagnetically ordered 
ground state has zero chirality, the finite energy skyrmion states 
are chiral doublets. It should be interesting to look for this
degeneracy in the excited states in numerical studies of finite
systems. This degeneracy is also reminiscent of the degeneracy
of the spinon states in the 1-d Haldane-Shastry model arising from
the Yangian symmetry. 

Our analysis goes through for any spin. In general for the
spin-S square lattice Heisenberg model we get deconfined 
`spin-S' spinons or the SU(2) vortices.

The spinon deconfinement that we have found has several interesting
consequences which we hope to discuss in the future: i) suggests a 
way to produce charge 2e skyrmions to get superconductivity in
the doped Mott insulator, ii) as the 
energy of a deconfined spinon $\approx \pi J$, its 
possible signature in the top of the spin wave band as well as
in the infrared and two magnon Raman measurements, iii)
short range RVB state viewed as the condensation of skyrmion and
antiskyrmion in the ground state; chiral symmetry broken 
Kalmayer-Laughlin like states as condensation of unequal density of 
skyrmion and antiskyrmion and iv) consequence of our present picture 
to skyrmion doping in quantum Hall ferromagnets.

I wish to thank R. Shankar for discussions and M.B. Silva Neto for
bringing reference (17) to my attention.

\end{multicols}

\begin{references}
\bibitem{note1}{ Even though equation (8) manifestly resembles 
a U(1) superfluid of a hard core bose fluid, actually it is 
an `SU(2) superfluid' in view of the presence of two independent 
goldstone modes}.
\end{references}
\end{document}